# INTERACTIONS BETWEEN THE MOLECULES OF DIFFERENT FULLERENES

## V.I.Zubov[1, 2, *]

[1]*Department of Theoretical Physics, Peoples' Friendship University, Moscow, Russia,*

[2] *Instituto de Física, Universidade Federal de Goiás, C.P. 131, 74001-970, Goiânia, GO, Brasil*

## ABSTRACT

Generalizing the procedures of Girifalco and of Verheijen *et al*. and using results of the preceding work it has been derived the interaction potential between the molecules of different fullerenes $C_n$ and $C_m$ at orientationally disordered (high-temperature solid and gaseous) phases. We have calculated the coefficients for interactions of the $C_{60}$ molecule with the molecules of higher fullerenes from $C_{70}$ to $C_{96}$ and with the smaller one, $C_{36}$. The dependence of the minimum point $r_o$ and of the depth of the potential well $\varepsilon$ on the numbers of the atoms in the molecules $m$ and $n$ has been investigated.

The theoretical studies of thermodynamic properties of orientationally disordered phases of fullerenes trace back to the work of Girifalco. [1] Considering that the $C_{60}$ molecule have near-spherical shape with the radius $a = 3.55 \cdot 10^{-8}$ cm he obtained the pair-wise intermolecular potential

$$\Phi_G \quad (r) = -\alpha\left(\frac{1}{s(s-1)^3} + \frac{1}{s(s+1)^3} - \frac{2}{s^4}\right) + \beta\left(\frac{1}{s(s-1)^9} + \frac{1}{s(s+1)^9} - \frac{2}{s^{10}}\right) \qquad (1)$$

where $s = r/2a$, $r$ is the distance between the centers of the molecules,

$$\alpha = n^2 A/12(2a)^6, \beta = n^2 B/90(2a)^{12}, \qquad (2)$$

---

• Correspondence: Department of Theoretical Physics, Ptoples' Friendship University, 117198, Moscow, Russia; Instituto de Física, Universidade Federal de Goiás, C.P. 131, 74001-970, Goiânia, GO, Brazil; E-mail: v_zubov@mail.ru, zubov@fis.ufg.br

$n = 60$ are the number of atoms in the $C_{60}$ molecule. The parameters $A$ and $B$ were fitted to experimental data for the lattice constant and heat of sublimation:

$$A = 3.200 \times 10^{-59} \text{ erg} \cdot \text{cm}^6, B = 5.577 \times 10^{-104} \text{ erg} \cdot \text{cm}^{12}. \tag{3}$$

The $C_{70}$ molecule has a form similar to an oblong uniaxial ellipsoid with semi-axes $a^{(1)} = a^{(2)} = 3.61 \cdot 10^{-8}$ cm and $a^{(3)} = 4.26 \cdot 10^{-8}$ cm.. Noticing that the it can be separated in five groups of 10 or 20 atoms, each one lying in a spherical shell of a certain radius $R_i$ ($1 \le i \le 5$), and generalizing the procedure of Girifalco,[1] Verheijen *et al*. [2] have obtained the pair-wise potential for orientationally disordered phases of this fullerene. It consists of a sum of 25 Girifalco-type (1) terms.

Kniaz, et al.[3] and Abramo and Caccamo [4] applied the Girifalco potential for the orientationally disordered modification of $C_{70}$. The radius of the approximating sphere $a$ was determined by fitting the calculated lattice constant to its experimental value. Early, the idea of a spherical approximation for some higher fullerenes with radius related to the number of atoms in the molecule was used by Saito *et al.* [5] and by Molchanov *et al.* [6] but regardless to the potentials.

A method was proposed for the calculation of the coefficients $\alpha$ and $\beta$ of the potential of Girifalco (1) basing on the spherical approximation of the form of the molecules, starting from their magnitudes for the $C_{60}$, [7] i.e. without additional fitting parameters. The effective radius of the molecule of arbitrary fullerene is related to that of the $C_{60}$ by

$$a_n = a_{60} \sqrt{n/60} \,. \tag{4}$$

Using this formula together with (2), it have been calculated the coefficients for a smaller and higher fullerenes, from the $C_{28}$ to the $C_{96}$. It is interesting that the coefficients $\alpha$ and $\beta$ decrease with increasing number of atoms in the molecule,[7] although the minimum point of the potential and the depth of its well of course increase. They have been utilized in the investigations of corresponding fullerites including their equilibrium with the gaseous phases.[8 – 10] Recently Fernandes *et al*. [11] have used the Girifalco potential (1) with our parameters[7] in computer simulations for $C_{70}$, $C_{76}$ and $C_{84}$ fullerites.

Great interest has been displayed also in investigations of mixtures of the fullerites. [12, 13] For the theoretical study of their thermodynamic properties it is necessary a knowledge of the intermolecular forces between the molecules of different fullerenes. Similarly to Girifalco[1] and Verheijen et al.,[2] averaging the atom-atom interactions (1) of two molecules $C_m$ and $C_n$ of effective radii $a_m$ and $a_n$ over all their orientations we obtain the intermolecular potential

$$\Phi_{mn}(r) = -\frac{\alpha}{s}\left(\frac{1}{(s-1)^3} + \frac{1}{(s+1)^3} - \frac{1}{(s-\delta)^3} - \frac{1}{(s+\delta)^3}\right)$$
$$+ \frac{\beta}{s}\left(\frac{1}{(s-1)^9} + \frac{1}{(s+1)^9} - \frac{1}{(s-\delta)^9} - \frac{1}{(s+\delta)^9}\right).$$

(5)

Here $s = r/D_{mn}$, $D_{mn} = (a_m + a_n)$, $a_m$ and $a_n$ are their effective radii, $\delta = (a_m - a_n)/(a_m + a_n)$, and the coefficients $\alpha$ and $\beta$ are defined by the formulae

$$\alpha = \frac{mnA}{48 a_m a_n (a_m + a_n)^4}, \quad \beta = \frac{mnB}{360 a_m a_n (a_m + a_n)^{10}}$$

(6)

One can readily see that when $m = n$, the potential (5) transforms to the Girifalco potential (1) for the two $C_n$ fullerene molecules. [7]

We have calculated the coefficients (6) for interactions of the $C_{60}$ molecule with the molecules of some higher fullerenes from $C_{70}$ to $C_{96}$ and with the smaller one, $C_{36}$. The parameters of this potential are given in Table 1, together with refined parameters of the Girifalco potential, and some potential curves are shown in Figs. 1 and 2.

It is apparent that the diameter of the hard core of the potential (5) $D_{mn}$ is defined as the arithmetic mean between those of $C_m - C_m$ and $C_n - C_n$ is used for the calculations of the coefficients $\alpha$ and $\beta$ (7). Formulae for the distance $r_0$ where $\Phi_{mn}(r_0) = 0$, for the minimum point $\sigma$ and for the depth of the potential well $\varepsilon$ are not available. It can be seen from Table 1 that the $r_0$ and $\sigma$ coincide with the corresponding arithmetic means within hundredths of percent while $\varepsilon$ with the geometric mean $\varepsilon_{mn} = \sqrt{\varepsilon_{mm} \varepsilon_{nn}}$ within tenths of percent.

Table I. Characteristics of the potentials (1) and (5) for different fullerenes

($D_{mn}$, $\sigma$ and $R_0$ in $10^{-8}$ cm)

| $M, n$ | $D_{mn}$ | $\delta$ | $\alpha (10^{-14}$erg) | $\beta (10^{-17}$ erg) | $\sigma$ | $R_0$ | $-\varepsilon/k_B$ (K) |
|--------|----------|----------|------------------------|------------------------|----------|-------|------------------------|
| 60, 60[a] | 7.100 | - | 7.494 | 13.595 | 9.599 | 10.056 | 3218 |
| 28, 28[b] | 4.850 | - | 16.058 | 286.64 | 7.358 | 7.808 | 1775 |
| 36, 36[b] | 5.500 | - | 12.49 | 104.90 | 8.000 | 8.460 | 2182 |
| 50, 50[b] | 6.481 | - | 8.980 | 28.19 | 8.976 | 9.443 | 2813 |
| 70, 70[b] | 7.669 | - | 6.424 | 8.338 | 10.161 | 10.622 | 3594 |
| 76, 76[b] | 7.991 | - | 5.916 | 5.281 | 10.480 | 10.946 | 3807 |
| 84, 84[b] | 8.401 | - | 5.353 | 3.539 | 10.880 | 11.358 | 4080 |
| 96, 96[b] | 8.981 | - | 4.684 | 2.074 | 11.468 | 11.936 | 4468 |
| 60, 36[c] | 6.300 | 0.127 | 9.365 | 34.813 | 8.800 | 9.260 | 2639 |
| 60, 70[c] | 7.384 | 0.0385 | 6.918 | 9.914 | 9.880 | 10.340 | 3400 |
| 60, 76[c] | 7.545 | 0.059 | 6.612 | 8.326 | 10.035 | 10.502 | 3497 |
| 60, 84[c] | 7.750 | 0.084 | 6.245 | 6.695 | 10.239 | 10.705 | 3616 |
| 60, 96[c] | 8.040 | 0.117 | 5.764 | 4.957 | 10.530 | 10.999 | 3777 |

[a] Girifalco [1]

[b] The work [7]

[c] The present work

The potential (5) with parameters listed in Table 1 can be utilized in computations of thermodynamic properties for the $C_{60}$ fullerite mixed with higher and smaller ones. One can use also the formulae (4) and (6) to calculate the parameters of the potential between other pairs of different fullerene molecules.

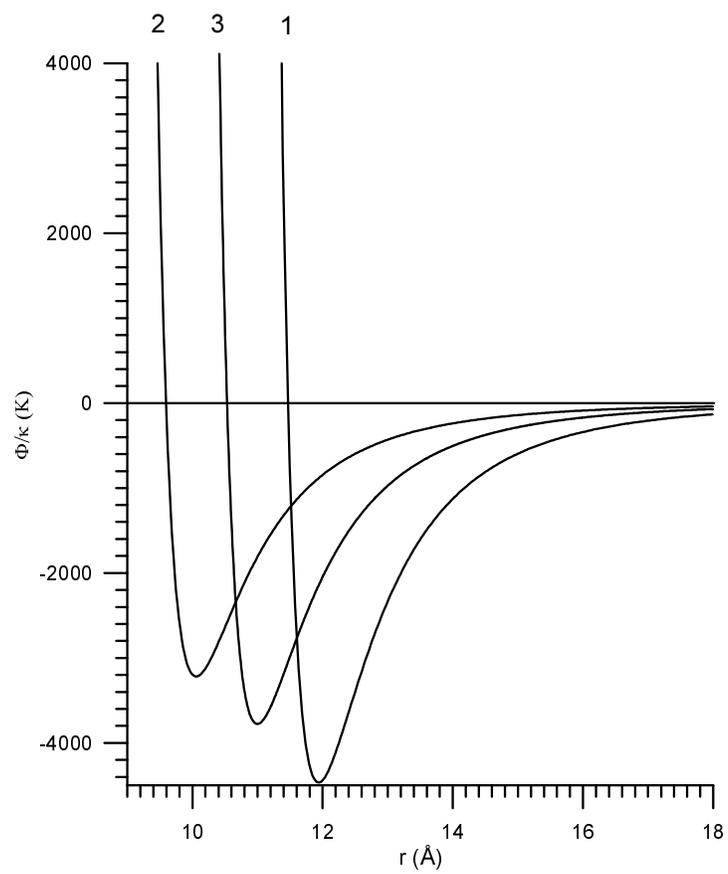

Fig. 1. Intermolecular potentials for $C_{60}$ - $C_{60}$ (1), $C_{96} - C_{96}$ (2) and $C_{60} - C_{96}$ (3).

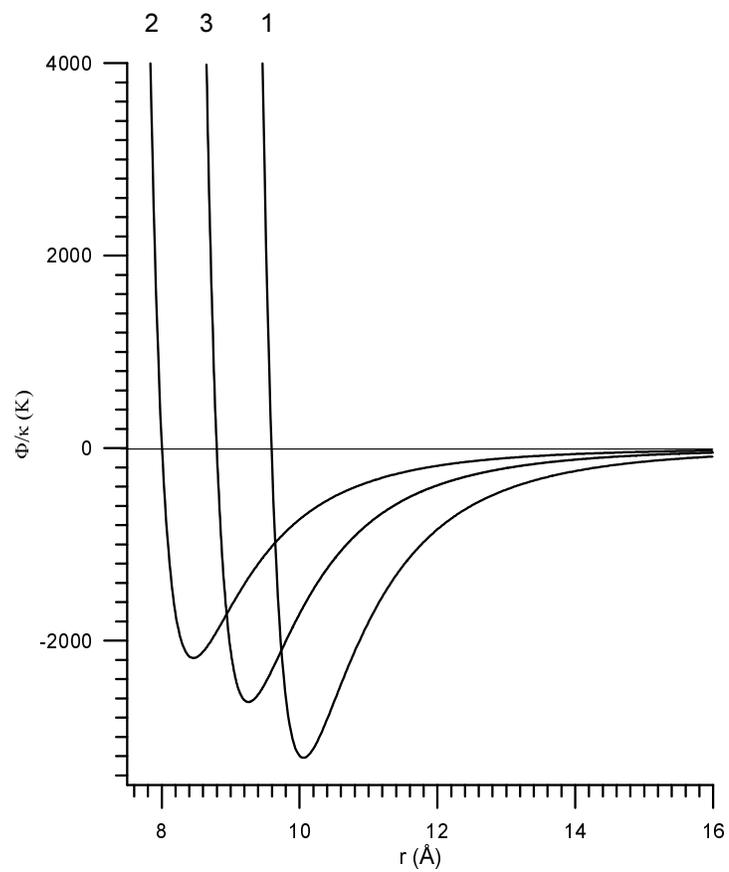

Fig. 2. Intermolecular potentials for $C_{60}$ - $C_{60}$ (1), $C_{36}$ - $C_{36}$ (2) and $C_{60}$ - $C_{36}$ (3).